\documentclass[preprint,showpacs,onecolumn,nofootinbib]{revtex4}
\pdfoutput=1
\usepackage[colorlinks=true,linkcolor=blue,urlcolor=blue,filecolor=black,citecolor=red,pdfstartview=FitV,pdftitle={},pdfsubject={},pdfkeywords={},pdfpagemode=None,bookmarksopen=true]{hyperref}
\usepackage{graphicx}
\usepackage{amsmath}
\usepackage{amsfonts}
\usepackage{amssymb,ulem}
\usepackage{dcolumn}

\usepackage{MnSymbol,wasysym}
\usepackage{braket}
\usepackage{color}%
\usepackage{eurosym}
\usepackage{calrsfs}
\usepackage[usenames,dvipsnames,svgnames]{xcolor}

\newcommand{\RNum}[1]{\uppercase\expandafter{\romannumeral #1\relax}}
\def\sideremark#1{\ifvmode\leavevmode\fi\vadjust{\vbox to0pt{\vss
 \hbox to 0pt{\hskip\hsize\hskip1em
 \vbox{\hsize1.5cm\tiny\raggedright\pretolerance10000
 \noindent #1\hfill}\hss}\vbox to8pt{\vfil}\vss}}}%

\usepackage[titletoc]{appendix}

\begin{document}
\baselineskip=0.6 cm
\title{Revisiting black hole thermodynamics in massive gravity: charged particle absorption and shell of dust falling}
\author{Shi-Qian Hu$ ^1$}
\email{mx120170256@yzu.edu.cn}
\author{Bo Liu$ ^{2,3,1}$}
\email{fenxiao2001@163.com}
\author{Xiao-Mei Kuang$ ^1$}
\email{xmeikuang@yzu.edu.cn}
\author{Rui-Hong Yue$ ^1$}
\email{rhyue@yzu.edu.cn}
\affiliation{$^1$ Center for Gravitation and Cosmology, College of Physical Science and Technology,
Yangzhou University, Yangzhou 225009, China}
\affiliation{$^2$ School of Physics, Northwest University, Xi'an, 710069, China}
\affiliation{$^3$ School of Arts and Sciences, Shaanxi University of Science and Technology, Xi'an, 710021, China}

\date{\today }
\begin{abstract}
\baselineskip=0.6 cm
We study the variation of the massive black hole  in both normal and extended thermodynamic phase spaces via two methods.  The first method includes a charged particle absorbed by the black hole, while the second method includes a shell of dust falling into the black hole. Using the former method, the first and second laws of thermodynamics are always satisfied in the normal phase space, while  in the extended phase space,  the first law of thermodynamics is also satisfied, but the validity of the second law of thermodynamics depends on the model parameters. Applying the latter method, the first and second laws of thermodynamics are both valid. We argue that the possible reason for the violation of the second law of thermodynamics via the former method may be attributed to the assumption that  the change of internal energy of the black hole
is equal to the energy of the particle. Finally, we show that the event horizon always exists to guarantee the validity of the weak cosmic censorship in both phase spaces, which means that the violation of the second law of thermodynamics under the assumption does not affect the weak cosmic censorship conjecture.
This further supports our argument that the assumption in the first method may be
the reason for the violation of the law and it requires deeper treatment.
\end{abstract}

\maketitle

\tableofcontents

\section{Introduction}
The energy in the form of Hawking radiation can be emitted from the horizon of black hole after one considers the quantum effect \cite{Bardeen:1973gs,Hawking:1975iha,Hawking:1976de}. Due to this, a black hole could be treated as an actual thermodynamic system, which is similar to the traditional thermodynamics\cite{Wallace:2017yfi}. This similarity leads to the definition of Bekenstein-Hawking entropy which is proportional to the area of the event horizon\cite{Bekenstein:1973ur,Bekenstein:1974ax,Bekenstein:1972tm,Christodoulou:1970wf}. The understanding of entropy production is a crucial point to quantify the black hole thermal systems. This is not only of the importance to comprehend the perception of the black hole physics, but also provides considerable new ideas to work on. Afterward, the relationship between gravitation, thermodynamics and quantum theory has been  attracting considerable attention.

Recently, one of progress in the development of black hole thermodynamics is to construct an extended phase space in black hole\cite{Kastor:2009wy,Dolan:2010ha,Cvetic:2010jb}. The idea is to treat the cosmological constant as the pressure of the black hole thermal system, and the thermodynamical volume is the thermodynamical conjugate of the pressure, i.e. $V=(\partial M/\partial P)|_{S,Q,J}$ \cite{Kubiznak:2014zwa,Kubiznak:2016qmn}. This volume is regarded as a volume covered by the event horizon of the black hole and the gravitational mass is regarded as enthalpy instead of energy as usual. In this framework, many remarkable properties of black holes have been studied, such as the Van der Waals behavior\cite{Kubiznak:2012wp}, solid/liquid phase transitions and triple points\cite{Altamirano:2013uqa}, reentrant phase transitions\cite{Altamirano:2013ane}, holographic heat engine\cite{Johnson:2014yja}, Joule-Thomson expansion\cite{Okcu:2016tgt} and so on. The subject has been generalized as black hole chemistry, see\cite{Kubiznak:2016qmn} as a review.
%
%
Based on the studies, the four laws of thermodynamics play an important role, therefore, it is interesting and important to explicitly visit their validity in the extended phase space of black hole thermodynamics.

 Besides, the validity of the first law does not always imply that the second law and the weak cosmic censorship conjecture(WCCC) are equally valid.
The weak cosmic censorship states that a singularity should be hidden by the horizons except at the Big Bang, so it is invisible to the observers outside. Cosmic censorship was proposed to avoid the breakdown of causality\cite{Penrose:1964wq,Penrose:1969pc,Hawking:1969sw} which is still a famous  open problem.  Thus, physicists keep interesting in the test of its validity, especially the validity of the WCCC. A common strategy is a particle absorbed by a black hole to see whether the horizon still exists for the validity of  WCCC, see for example\cite{Wald:1974rm} and the extension to other test fields\cite{Semiz:2005gs,Duztas:2013wua,Duztas:2013gza}. Afterward, the WCCC has been studied widely in various black holes through particle absorption or scattering of the test field\cite{Hubeny:1998ga,Jacobson:2009kt,Barausse:2010ka,Pani:2010jz,Sorce:2017dst,Matsas:2007bj,Duztas:2016xfg}. The previous studies show that the validity of WCCC depends on both the state of black hole and the method of perturbation, so there is no general way to discuss the conjecture.

More recently,
Gwak investigated the variation of horizon and the laws of thermodynamics under charged particle absorption in the extended phase space,  and also checked whether the WCCC is still valid\cite{Gwak:2017kkt}. It was shown that the second law of thermodynamics is violated with the contribution of $PV$ term while the WCCC is still valid. This work was soon extended to other gravitational theories\cite{Gim:2018axz,Chen:2018yah,Zeng:2019jta,Gwak:2019asi,Han:2019kjr,Chen:2019pdj,Han:2019lfs,Wang:2019dzl,Zeng:2019aao,Zeng:2019hux,Hong:2019yiz,Mu:2019bim,He:2019kws}. Now we will focus this work in massive gravity which is one example of a theory that modifies gravity at large distance-scales.

As is known that black holes are important theoretical tools for exploring general gravity as they provide a particle environment for testing  gravity. Moreover,
we know that the graviton is massless in Einstein's theory of general relativity. Recently, some cosmologists proposed the idea of massive graviton to modify general relativity to see whether it's a conspiracy theory. The first attempt of the construction in massive gravity was done by Fierz and Pauli, but their construction can not recover general relativity with the limit of $m_{graviton}=0$\cite{Fierz:1939ix}. Afterward, Vainshtein proposed a nonlinear model for massive gravity \cite{Vainshtein:1972sx} but it suffers
Boulware-Deser(BD) ghosts \cite{Isham:1971gm,Boulware:1973my}. Recently, more non-linear terms have been introduced into massive gravity, which avoids this instability \cite{deRham:2010ik,deRham:2010kj,Hassan:2011hr,Hassan:2011vm,Hassan:2011tf,Desai:2010ea,Ludeling:2012cu,Deser:2014fta}. Various phenomenology of massive gravity has  been widely investigated. Especially,  its thermodynamical phase transition and the related studies in extended phase space  have been  addressed in \cite{Cai:2014znn,Xu:2015rfa,Zou:2017juz,Cisterna:2018jqg,Hendi:2017fxp,Liu:2018jld,Mo:2017nes,Fang:2017nse,Hu:2019wwn,Hendi:2018xuy,Nam:2019zyk}.

In this paper, we shall revisit the thermodynamics and WCCC in four dimensional massive gravity with a negative cosmological constant\cite{Vegh:2013sk,Cai:2014znn}. We first use Gwak's proposal by considering a charged particle absorbed by the black hole,  then we include a shell of dust falling into the black hole as an alternative method.

The outline of the paper is as follows. In section \ref{sec:massive gravity}, we briefly review the AdS black hole solution in massive gravity. In section \ref{sec:particle absorbed}, we take into account the study of the thermodynamics of massive gravity black hole by absorbing a charged particle both in the normal and extended phase space. While in section \ref{sec:shell}, we investigate the laws of thermodynamics under a shell of infalling dust. Finally, in section \ref{sec:wcc}, we concentrate on the weak cosmic censorship conjecture under the charged particle absorbing which shows the horizon always exists. Conclusions and discussions are shown in the last section.

\section{Black holes with spherical horizon in massive gravity}\label{sec:massive gravity}
The action of the four-dimensional massive gravity we are considering is given by (in the units $c=G=\hbar=k_B=4\pi \varepsilon_0= 1$)\cite{Vegh:2013sk}
\begin{equation}
S=\frac{1}{16\pi}\int
d^{4}x\sqrt{-g}\left[R+\frac{6}{l^2}-\frac{1}{4}F^2+m_g^2\sum_{i=1}^4c_i\mathcal{U}_i(g,f)\right]\label{1}
\end{equation}
where $m_g$ is the mass of graviton in the theory.
In the action, the last term represents massive potentials associated with the graviton which breaks the diffeomorphism invariance in the bulk producing momentum relaxation in the dual boundary theory.
The couplings parameters $c_i$ are constants, while $f$  denotes the reference metric, and $\mathcal{U}_i$  are symmetric polynomials of the eigenvalue of the $4$ $\times$ $4$ matrix $\mathcal{K}^{\mu}_{~\nu}\equiv\sqrt{g^{\mu\alpha}f_{\alpha\nu}}$ with the forms
\begin{eqnarray}
\mathcal{U}_1&=&[\mathcal{K}],~~~\mathcal{U}_2=[\mathcal{K}]^2-[\mathcal{K}^2]\nonumber,\\
\mathcal{U}_3&=&[\mathcal{K}]^3-3[\mathcal{K}][\mathcal{K}^2]+2[\mathcal{K}^3]\nonumber,\\
\mathcal{U}_4&=&[\mathcal{K}]^4-6[\mathcal{K}^2][\mathcal{K}]^2+8[\mathcal{K}^3][\mathcal{K}]+3[\mathcal{K}^2]^2-6[\mathcal{K}^4], \label{2}
\end{eqnarray}
where $[\mathcal{K}]=\mathcal{K}^{\mu}_{~\mu}$ and the square root in $\mathcal{K}$ is defined as $(\sqrt{\mathcal{K}})^{\mu}_{~\nu}(\sqrt{\mathcal{K}})^{\nu}_{~\lambda}=\mathcal{K}^{\mu}_{~\lambda}$.

The static spherical black hole solution of the above action yields~\cite{Cai:2014znn}
\begin{equation}
ds^2=-f(r)dt^2+\frac{dr^2}{f(r)}+r^2(d\theta^2+\sin^2\theta d\varphi^2). \label{3}
\end{equation}
 Following \cite{Vegh:2013sk}, we take the reference metric as $f_{\mu\nu}=\mathrm{diag}(0,0,c_0^2h_{ij})$.
Then with the ansatz \eqref{3} and the reference metric, we obtain $\mathcal{U}_1=2c_0/r$, $\mathcal{U}_2=2c_0^2/r^2$, and $\mathcal{U}_3=\mathcal{U}_4=0$, thus the solutions of fields are
\begin{eqnarray}
&&F_{\mu\nu}=\partial_{\mu}A_{\nu}-\partial_{\nu}A_{\mu},~~~A=-\frac{q}{r} dt,\\
&&f(r)=1+\frac{r^2}{l^2}-\frac{2\mathcal{M}}{r}+\frac{q^2}{r^{2}}+\frac{c_0c_1m_g^2}{2}r+c_0^2c_2m_g^2.\label{6}
\end{eqnarray}
According to the Hamiltonian approach, the parameters $\mathcal{M}$ and $q$ are related to the mass and charge of the black hole as\cite{Cai:2014znn}
\begin{eqnarray}
M=\frac{V_2}{4\pi}\mathcal{M}, ~~~~~~~~~~~~~~Q=\frac{V_2}{4\pi}q \label{7}
\end{eqnarray}
where $V_2$ is the volume of the two dimension space.
 Using $f(r)|_{r=r_h} =0$, where $r_h$ denotes the radius of horizon, we  derive the mass as
\begin{equation}
M=\frac{r_h}{2}+\frac{q^2}{2r_h}+\frac{1}{2}c_0^2 c_2 m_g^2 r_h +\frac{1}{4}c_0 c_1 m_g^2 r_h^2+\frac{r_h^3}{2 l^2} . \label{8}
\end{equation}
The Hawking temperature and the entropy of the black hole are
\begin{eqnarray}
T&=&\frac{f'(r_h)}{4\pi}=\frac{\ c_0 c_1 m_g^2}{4\pi}+\frac{1}{4\pi r_h} +\frac{c_0^2 c_2 m_g^2}{4 \pi r_h}-\frac{q^2}{4\pi r_h^3}+\frac{3r_h}{4\pi l^2},  \label{9} \\
S&=&\int_0^{r_h} dr \frac{1}{T}\left(\frac{\partial M}{\partial
r}\right)_{q,l,c_i}=\pi r_h^2, \label{10}
\end{eqnarray}
 respectively.

Thermodynamic and the related studies of massive black hole have been done in \cite{Cai:2014znn,Xu:2015rfa,Zou:2017juz,Cisterna:2018jqg,Hendi:2017fxp,Liu:2018jld,Mo:2017nes,Fang:2017nse,Hu:2019wwn,Hendi:2018xuy,Nam:2019zyk}. Thus, here as a revisiting, we will study the thermodynamics of massive gravity black hole  via two methods: one is the black hole  absorbs a charged particle, and the other is a shell of dust falls into the black hole. We will mainly focus on the validity of the thermodynamical laws  both in  normal phase space and extended phase space.

\section{Thermodynamics under charged particle absorption}\label{sec:particle absorbed}
In this section, we will study the thermodynamics of massive gravity black hole  by absorbing a charged particle.

\subsection{Energy and momentum of a particle absorbed by the black hole}

Upon a charged particle is absorbed, the charged massive gravity black hole will be perturbed so that the transferred energy, charge and  angular momentum derivate from the energy, charge and angular momentum fluxes of the particle at the outer horizon. The Hamiltonian of the charged particle is
\begin{equation}
\mathcal{H}=\frac{1}{2}g^{\mu\nu}(P_\mu-e A_\mu)(P_\nu-e A_\nu) \label{15}
\end{equation}
and the Hamiltonian-Jacobi action is
 \begin{equation}
S=\frac{1}{2}m^2 \lambda-E t+L\phi+S_r(r)+S_\theta(\theta), \label{16}
\end{equation}
where the four-momentum $P_\mu$ of the particle is defined as $P_\mu=\partial_\mu S$. Here $m^2$, $e$ and $\lambda$ are  the mass, charge of the particle and an affine parameter, respectively.  Due to the symmetry of the metric \eqref{3}, the conserved quantities $E$  and $L$  with respect to $t$ and  $\varphi$ are the energy and angular momentum of the particle, respectively.
To proceed, we combine \eqref{3} and \eqref{15}-\eqref{16} to deduce
\begin{equation}
m^2-f(r)^{-1}(-E-e A_t)^2+f(r)\Big(\partial_{r}S_r(r)\Big)^2+r^{-2}\Big((\partial_{\theta}S_\theta)^2+\sin^{-2}{\theta}L^2\Big)=0. \label{18}
\end{equation}
The angular part of the above expression can be defined separately as
\begin{eqnarray}
K&\equiv&\Big(\partial_{\theta}S_\theta(\theta)\Big)^2+\sin^{-2}{\theta}L^2
=-m^2 r^2+\frac{r^2}{f(r)}(-E-e A_t)^2-r^2 f(r)\Big(\partial_{r}S_r(r)\Big)^2. \label{19}
\end{eqnarray}
Thus, \eqref{16} is rewritten as
\begin{equation}
S=-E t+L\phi+\int{dr}\sqrt{R}+\int{d\theta}\sqrt{\Theta} \label{HJ}
\end{equation}
where
\begin{equation}
S_r\equiv\int{dr}\sqrt{R},~~S_\theta\equiv\int{d\theta}\sqrt{\Theta},~~\Theta\equiv K-\sin^{-2}{\theta}L^2,
\end{equation}
\begin{equation}
R=\frac{1}{r^2f(r)}(-K^2-u^2r^2)+\frac{1}{r^2f(r)}\left(\frac{r^2}{f(r)}(-E-e A_t)^2 \right).
\end{equation}
Subsequently, the radial momentum $P^r$ and angular momentum $P^\theta$ can be computed
\begin{eqnarray}
P^r&=&g^{rr}\partial_{r}S_r(r)=f(r)\sqrt{\frac{-K-m^2 r^2}{r^2 f(r)}+\frac{(-E-eA_t)^2}{f^2(r)}}, \label{21} \\
P^\theta&=&g^{\theta\theta}\partial_{\theta}S_\theta(\theta)=
\frac{1}{r^2}\sqrt{K-\frac{1}{\sin^2{\theta}}L^2}.\label{22}
\end{eqnarray}

Following \cite{Gwak:2017kkt}, we assume that the charged particle is  completely absorbed by the black hole when it passes through  the outer horizon.  Subsequently, it is impossible
for an observer outside the horizon to distinguish the conserved quantities of the particle from those of the black hole. Especially, a connection between conserved quantities and momenta at any radius position can be derived. Then the limit  \eqref{21} at the outer horizon will give us the following relation between the conserved quantities and the radial momentum
\begin{equation}
E=\frac{q}{r_h} e+|P^r|.\label{23}
\end{equation}
The equation above implies that both momentum and electric charge of the particle contributes to the energy. We note that $q/r_h$ is the electromagnetic potential at the event horizon
and $|P^r|$ should be positive.
 When electrical attraction acts on the particles, the total energy of the particles can be negative. But we simply choose the signs in front of $E$ and $|P^r|$ to be both positive, because when the particle falls into the black hole, a positive value of the energy must be satisfied \cite{Gwak:2017kkt,Zeng:2019jrh,Gwak:2019asi,Zeng:2019jta}.

 The relation involves interaction between the particle and the black hole, and we have assumed there is no energy loss in the process.
  Namely, the charge of the particle $e$ is equal to the varied charge of the black hole $dQ$.
The energy of the particle is expressed in terms of $e$ and $|P^r|$ near the horizon in \eqref{23}. It is crucial for us to find a corresponding thermodynamical term that also contains the variation of $e$ and $|P^r|$. Following Gwak \cite{Gwak:2017kkt}, we also assume that the energy of the particle changes the internal energy of the black hole as a charged particle is swallowed by the black hole.

\subsection{Thermodynamics in the normal phase space}
We will first study the validity of  the thermodynamical laws of the massive black hole in normal phase space. Since in the normal phase space, the mass of black hole represents the internal energy, so with the assumption of $E=dU$, \eqref{23} can be rewritten as
\begin{equation}
E=dU=dM=\frac{q}{r_h}e+ |P^r|.\label{49}
\end{equation}
To analyze the first law under the particle absorption, we rewrite  the variation of the entropy from the Bekenstein's area law as $dS_h=2\pi r_h dr_h$.
The variation of the event horizon $dr_h$  determined by  the charge, energy, and radial momentum of the absorbed particle will directly contribute to the changes of the function $f(r)$.  Then, the change of  $f(r)$ could lead to the moved event horizon $r_h+dr_h$, even so, $f(r_h)$ near event horizon will not change because of $f(r_h+dr_h)=0$, then we have  $df(r_h)=df_h=0$, i.e.,
\begin{equation}
df_h=\frac{\partial{f_h}}{\partial{M}}dM+\frac{\partial{f_h}}{\partial{Q}}dQ+
\frac{\partial{f_h}}{\partial{r_h}}dr_h=0.\label{50}
\end{equation}
With the help of \eqref{49} and \eqref{50}, we can obtain  $dr_h$ and thus, the  $dS_h$ is
\begin{equation}
dS_h=\frac{24 \pi |P^r| r^3_h}{3 c_0 c_1 m_g^2 r^3_h+12 M r_h+12  r^4_h/l^2-12 q^2}.\label{51}
\end{equation}
Combining \eqref{9} and \eqref{51} gives us the relation $TdS_h=|P^r|$. Then \eqref{23} becomes $E=e~q/r_h +TdS_h$. Subsequently, due to  $E=dM$ and $e=dQ$, we obtain the form of the variation
\begin{equation}
dM=TdS_h+\Phi dQ
\end{equation}
which is the first law of thermodynamics of massive black hole in normal phase space.

We then employ \eqref{51} to check the validity of the second law of the thermodynamics, which states that the entropy has to increase after a particle absorption.

Firstly, for the case of the extremal massive black hole which means that the temperature vanishes, $T=0$ in \eqref{9} gives us the marginal mass
\begin{equation}
M_e=r_e+c_0^2 c_2 m_g^2 r_e+\frac{3}{4}c_0 c_1 m_g^2 r_e^2+2 r_e^3/l^2\label{35}
\end{equation}
where $r_e$ is the horizon of the extremal black hole. It is straightforward to reduce $r_e$ as a function of $q, l, c_0, c_1$ and $c_2$, which we don't show due to the complexity. Substituting \eqref{35} into \eqref{51}, we obtain that $dS_h=\infty$ holds for the extremal black hole.

\begin{figure}
   \centering
\includegraphics[width=8cm]{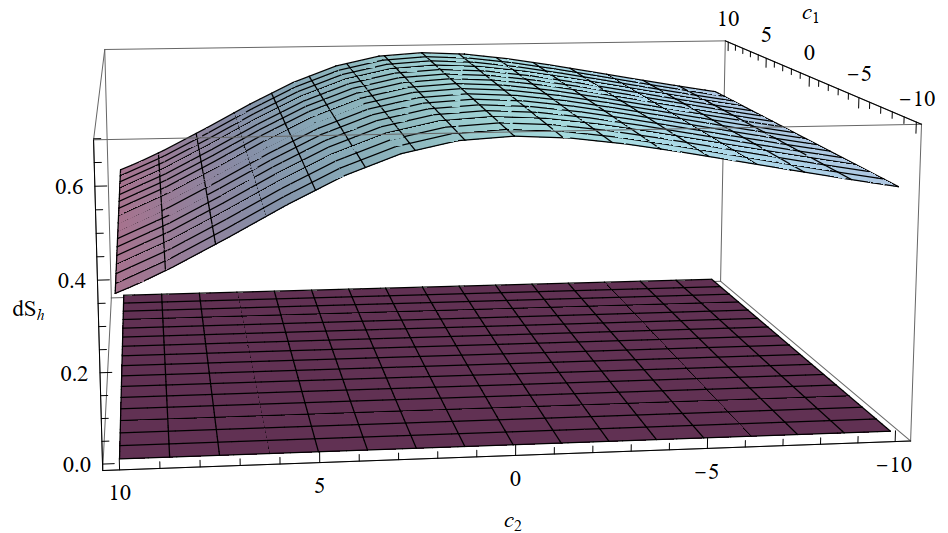}

\caption{ $dS_h$ near the horizon with $c_1,c_2$ for non-extremal black hole in normal phase space.}   \label{fig4}
\end{figure}

\begin{figure}

\includegraphics[width=6cm]{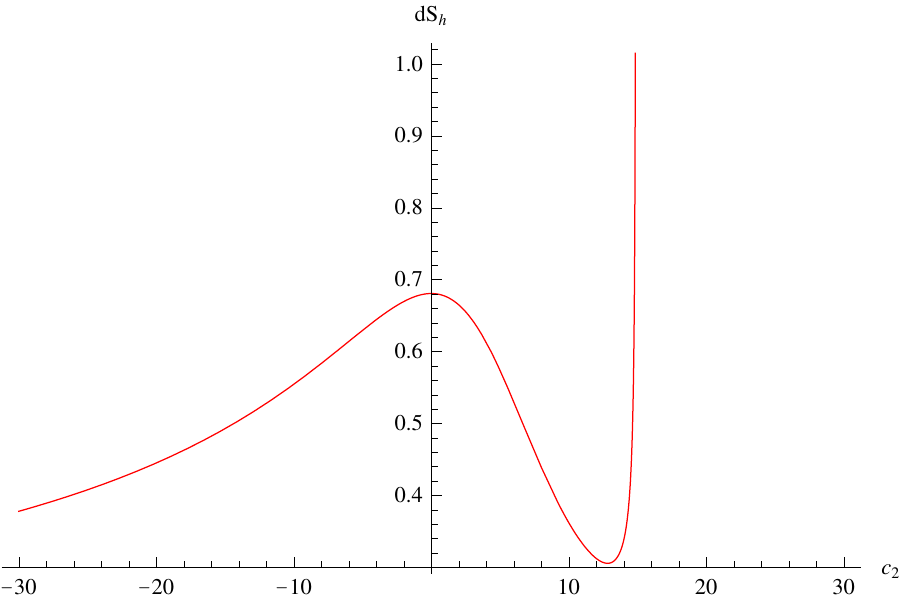}~~~~~~
\includegraphics[width=6cm]{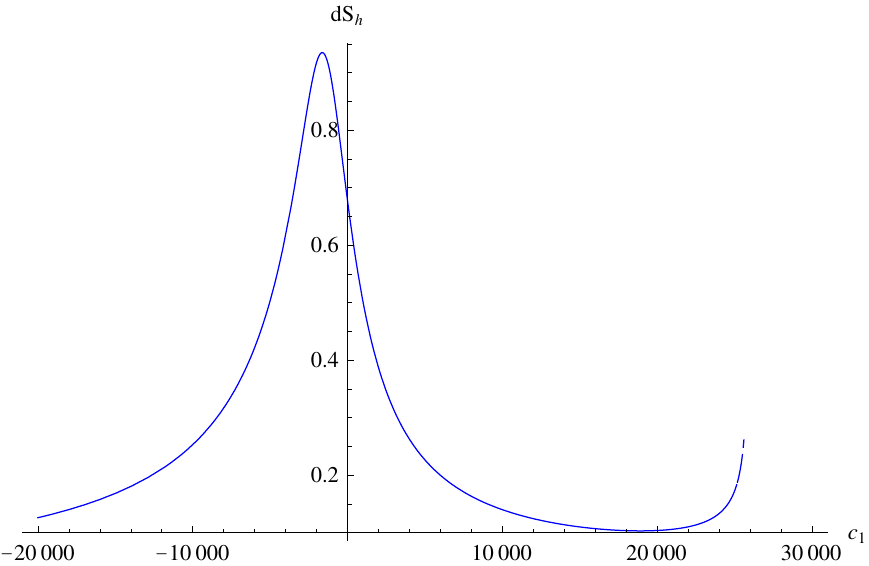}
\caption{Left:The relation between $dS_h$ and $c_2$ with fixed $c_1=5$ ;
Right:The relation between $dS_h$ and $c_1$ with fixed $c_2=5$.}   \label{fig5}
\end{figure}

In the case of the non-extremal black hole, $dS_h$ is always positive for any $c_i(i=1,2)$.
We plot the relation between $dS_h$ and samples of  $c_i$ in Fig.\ref{fig4}. It is obvious that the value of $dS_h$ is always larger than zero. More explicit relations between $dS_h$  and $c_1,c_2$ are shown in Fig.\ref{fig5}.

The above study shows that the variation of entropy is always positive in the normal phase space. That is to say, the second law of thermodynamics is valid in the normal phase space under the particle absorption.  We turn to the case with $PV$ term in the next subsection.

\subsection{Thermodynamics in the extended phase space}
The PV criticality  characterizing by treating   the pressure as $P=3/8\pi l^2$ has been studied in \cite{Xu:2015rfa,Zou:2017juz,Cisterna:2018jqg}. Here we first briefly review the main properties in the extended phase space and then check the validity of the laws of thermodynamics via absorbing a charged particle.

In  the extended phase space, the corresponding conjugate quantity of the pressure  $P=3/8\pi l^2$ is treated as volume $V$ \cite{Dolan:2010ha,Kastor:2009wy}. Then the mass \eqref{8}
and the temperature \eqref{9} are  rewritten, respectively, as
\begin{eqnarray}
M&=&\frac{r_h}{2}+\frac{q^2}{2r_h}+\frac{1}{2}c_0^2 c_2 m_g^2 r_h +\frac{1}{4}c_0 c_1 m_g^2 r_h^2+\frac{4}{3}P \pi r_h^3 ,\label{eq:M2}\\
T&=&\frac{\ c_0 c_1 m_g^2}{4\pi}+\frac{1}{4\pi r_h} +\frac{c_0^2 c_2 m_g^2}{4 \pi r_h}-\frac{q^2}{4\pi r_h^3}+2 P r_h, ~~~~S=\pi r_h^2. \label{eq:T2}
\end{eqnarray}
The other conjugate quantities of the intensive paraments $P,Q,c_1$ and $c_2$ are
\begin{equation}
V=\frac{\partial M}{\partial P}=\frac{4}{3}\pi r_h^3, ~~~\Phi=\frac{\partial M}{\partial Q}=\frac{q}{r_h},~~A=\frac{\partial M}{\partial c_1}=\frac{c_0 m_g^2 r_h^2}{4},~~B=\frac{\partial M}{\partial c_2}=\frac{c_0^2 m_g^2 r_h}{2},\label{11}
\end{equation}
where one has to consider the coupling constants $c_i$ as thermodynamical variables.  All the thermodynamical quantities satisfy the first law of thermodynamics
 \begin{equation}
 dM=TdS+\Phi dQ+V dP+A dc_1+B dc_2,\label{12}
\end{equation}
and the generalized Smarr relation is
 \begin{equation}
M=2TS-2VP+\Phi Q-A c_1.\label{13}
\end{equation}
We note that in the extended phase space,  $M$ plays the role of enthalpy rather than internal energy of the thermodynamical system, i.e, we have
\begin{equation}
M=U+VP+A c_1+B c_2.\label{14}
\end{equation}

In this case, when the fluxes of the energy come into the event horizon, it will change the internal energy of the black hole which can be given as a function of $U(Q,S,V,c_1,c_2)$. From \eqref{14}, the conversed energy and charge of the particle are
\begin{equation}
E=dU=d(M-PV-A c_1-B c_2),~~~~~e=dQ.\label{31}
\end{equation}
Consequently, the energy relation near the horizon in \eqref{23} can be expressed as
\begin{equation}
dU=\frac{q}{r_h} dQ+|P^r|.\label{25}
\end{equation}
Similar to the analysis in the last subsection,  the slight variations of the redshift function is
\begin{equation}
df_h=\frac{\partial{f_h}}{\partial{M}}dM+\frac{\partial{f_h}}{\partial{Q}}dQ+\frac{\partial{f_h}}{\partial{P}}dP+
\frac{\partial{f_h}}{\partial{r_h}}dr_h+\frac{\partial{f_h}}{\partial{c_1}}dc_1+\frac{\partial{f_h}}{\partial{c_2}}dc_2=0.\label{27}
\end{equation}
Combining the \eqref{14} and \eqref{27}, the contribution of $dP,dM,dc_1$ and $dc_2$ terms can be eliminated directly, after which the above expression gives us
\begin{equation}
dr_h=\frac{12(c_1 dA+c_2 dB+|P^r|) r_h}{-12M +r_h(12 +12c_0^2 c_2 m_g^2 +9 c_0 c_1 m_g^2 r_h+16 P\pi r_h^2)}.\label{28}
\end{equation}
Subsequently, we obtain the change of the entropy and volume
\begin{eqnarray}
dS&=&\frac{24\pi(c_1 dA+c_2 dB+|P^r|) r_h^2}{-12M +r_h(12 +12c_0^2 c_2 m_g^2 +9 c_0 c_1 m_g^2 r_h+16 P\pi r_h^2)},\label{29}\\
dV&=&\frac{16\pi(c_1 dA+c_2 dB+|P^r|) r_h^3}{-12M +r_h(12 +12c_0^2 c_2 m_g^2 +9 c_0 c_1 m_g^2 r_h+16 P\pi r_h^2)}.\label{30}
\end{eqnarray}
From the  above formulas and the related thermodynamical quantities, we obtain the relation
 \begin{equation}
 T dS-P dV-c_1 dA-c_2 dB=|P^r|.
 \end{equation}
Thus, the expression of the internal energy in \eqref{31} becomes
\begin{equation}
dU=\Phi dQ+T dS-P dV-c_1 dA-c_2 dB\label{32}
\end{equation}
and then combining \eqref{14} and \eqref{32} gives
\begin{equation}
dM=TdS+\Phi dQ+V dP+A dc_1+B dc_2,\label{33}
\end{equation}
which is nothing but the first law of thermodynamics \eqref{12} of massive black hole in the extended phase transition. In other words, the first law of thermodynamics holds in the extended phase space under the particle absorption.

The next step is to analyze the sign of \eqref{29} to see if the second law is valid in the extended phase space.

\begin{figure}
  \centering
\includegraphics[width=6cm]{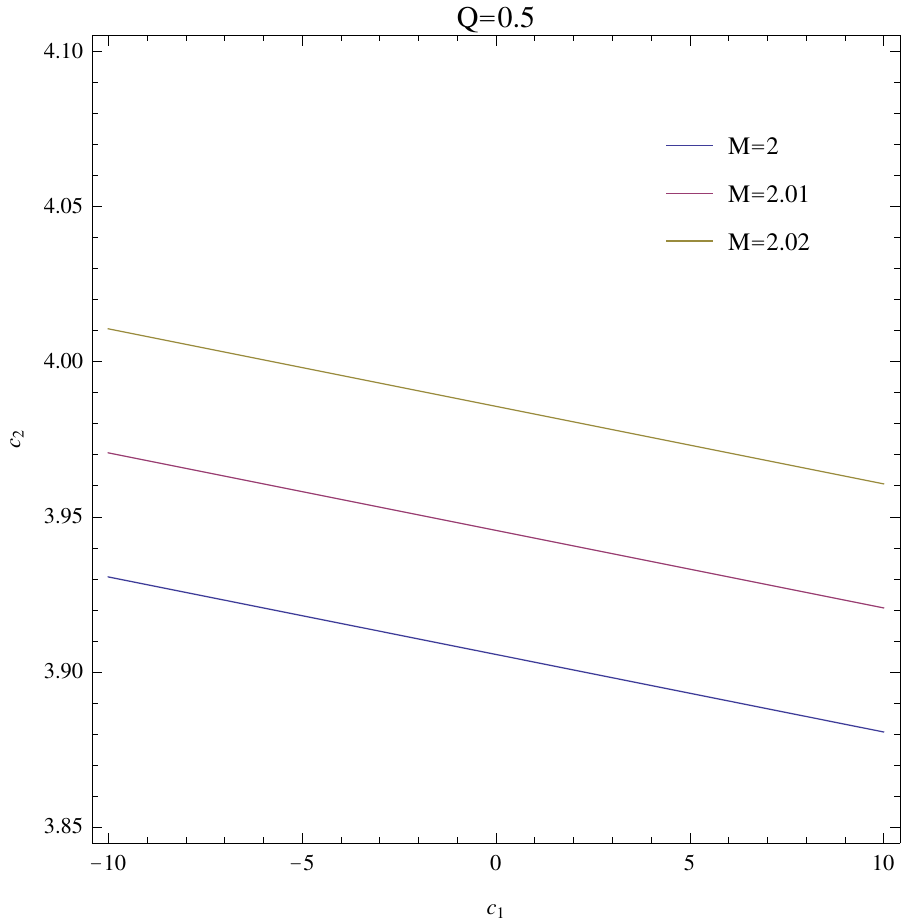}\hspace{1cm}
\includegraphics[width=6cm]{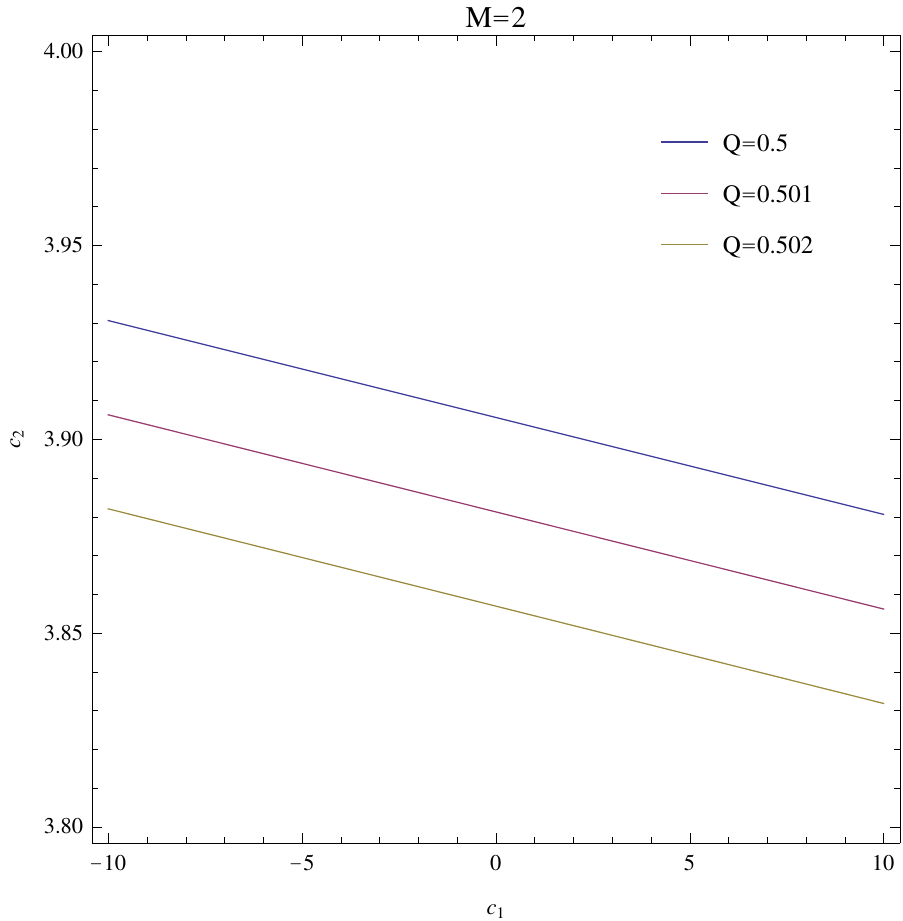}
\caption{The diagrams of critical  $c_1$ and  $c_2$ for different $M$(left) and different $Q$(right).}   \label{fig1}
\end{figure}

Taking the derivative of $A$ and $B$ in respective to $r_h$ in \eqref{11} gives us  $dA=c_0 m_g^2 r_h/2 dr_h, dB=c_0^2 m_g^2/2 dr_h$, and then substituting them into \eqref{29} gives us
 \begin{equation}
dS=\frac{24\pi|P^r| r_h^2}{-12M+r_h(12+6c_0^2 c_2 m_g^2+3c_0 c_1 m_g^2 r_h+16 P\pi r_h^2)}.\label{34}\\
 \end{equation}
Since the numerator of the above $dS$ is always  positive, we can only focus on the sign of the denominator of \eqref{34}, which will be labeled as $\Delta S_1$.  Apparently, the value of $\Delta S_1$ depends on the model parameters $c_1$ and $ c_2$. As an attempt, we set $c_0=100,m_g=0.01,p=1,P^r=1$ and  list the value of $\Delta S_1$  for different $c_2$ with fixed $c_1=5$ in table \ref{table1},
\begin{table}[h]
\begin{center}
\begin{tabular}{|c|c|c|c|c|c|}
 \hline
$M$&$Q$&$c_1$&$c_2$&$r_h$&$\Delta S_1$ \\ \hline
$2$&0.5&5&3.6931&0.5099&0.1171\\ \hline
$2$&0.5&5&3.7931&0.5049&0.0587\\ \hline
$2$&0.5&5&3.8931&0.5&$2.88058\times10^{-6}$ \\ \hline
$2$&0.5&5&3.9931&0.4951&-0.0589\\ \hline
$2$&0.5&5&4.0931&0.4903&-0.1181\\ \hline
 \end{tabular}
 \caption{\label{table1}Numerical results of  $\Delta S_1$  for different $c_2$ with fixed $c_1=5$.}
 \end{center}
\end{table}
from which there exists  certain $c_2$ making $\Delta S_1$ turn from positive to negative as it increases. Moreover, in Fig.\ref{fig1},
we plot the boundary lines  $(c_{1c},c_{2c})$ for which $\Delta S_1\to 0$. So for the parameters below the lines $\Delta S_1$ is always positive which denotes that
the second law of thermodynamics is saved, while for the parameters above each line, one always has  $\Delta S_1< 0$ which implies the violation of the second law. It is also obvious from the figure that
the borderline depends on $M$ and $Q$, and the  parameter range holding the second law is narrower for smaller $M$ but bigger $Q$.

\begin{figure}
   \centering
\includegraphics[width=8cm]{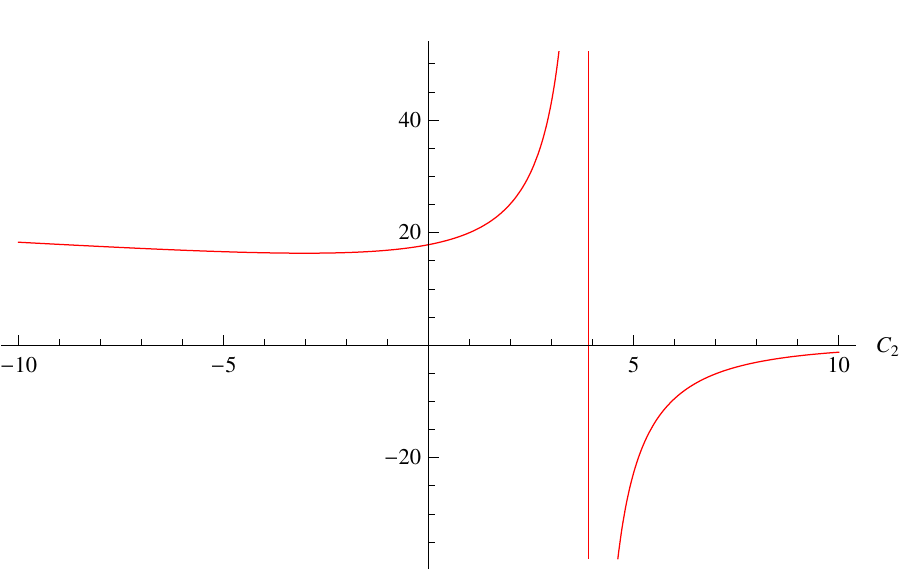}
\caption{$dS$ with $M=2, Q=0.5, c_1=5$ near the critical point $c_{2c}$.}   \label{fig3}
\end{figure}

It is worthwhile to point out that  $\Delta S_1=0$ means the change of entropy $dS$ is divergent at the points $(c_{1c},c_{2c})$ on the critical line. That's to say,
for a given value of $c_1$, there is a certain $c_{2c}$ that leads to the divergence of $dS$ as shown in Fig.\ref{fig3} . But we do not need to worry about this point because the importance is  the change of the sign in $dS$. And the degree of freedom  will go to maximal as $c_2$ approaches to $c_{2c}$ which is denoted by the vertical line in the figures. Actually, the critical point of $c_2$ is non-physical that the black hole will not reach this point because of the violation of the second law of thermodynamics.

Before closing this section, let us consider the case of the extremal massive black hole. We substitute the marginal mass  \eqref{35}  into \eqref{34},  then we could obtain
\begin{equation}
dS_{extremal}=-\frac{4\pi |P^r| r_e}{c_0 c_1 m_g^2 r_e+c_0^2 c_2 m_g^2+8\pi P r_e^2}.\label{37}
\end{equation}
Since the denominator of $dS_{extremal}$ is always positive,  we follow the strategy in the non-extremal case and  define the numerator as $\Delta S_2$.  For fixed $c_0=100,m_g=0.01,p=1, q=1/2$.
We list the numerical result of $\Delta S_2$ in table \ref{table2}. The result is similar to the case above, which states that  the violation of the second law of thermodynamics depends on the model parameters.

\begin{table}[h]
\begin{center}
\begin{tabular}{|c|c|c|c|c|c|}
 \hline
$c_1$&$c_2$&$r_e$&$\Delta S_2$ \\ \hline
$-2$&-6.07319&0.493109&-0.0833\\ \hline
$-2$&-6.17319&0.49656&-0.0414\\ \hline
$-2$&-6.27319&0.5&$1.93433\times10^{-6}$ \\ \hline
$-2$&-6.37319&0.50343&0.0410\\ \hline
$-2$&-6.47319&0.506849&0.0816\\ \hline
 \end{tabular}
 \caption{\label{table2}Numerical results of $\Delta S_2$ for different $c_2$ with fixed $c_1=-2$ in extremal case.}
 \end{center}
\end{table}


Now we briefly  summarize the results by absorption a charged particle in massive black hole: \emph{(i)} Both the first law and second law of thermodynamics hold in normal phase space. \emph{(ii) }When we consider the extended thermodynamics by involving $PV$ term, the first law is always saved but the second law can be violated depending on the model parameters.
This violation may be brought in by  the assumption of the internal energy of the black hole is equal to the energy of the particle.

In the next section, as an alternative strategy, we will consider a thin shell falling into the massive black hole instead of particle absorption.
Then we calculate the mass of shell when it arrives at the horizon of the black hole. We revisit the validity  of the laws of thermodynamics in both extended phase space and normal phase space.
\section{Thermodynamics under a shell of infalling dust}\label{sec:shell}
Considering a thin shell approaching to the  horizon and using the technique presented in appendix A, we obtain the
  external of the shell in terms of  $(M,Q,P,r,c_1,c_2)$
\begin{eqnarray}
n_{r}^+&=&\frac{\displaystyle\left(1+\frac{8\pi P}{3}r^2-\frac{2\mathcal{m}}{r}+\frac{Q^2}{r^{2}}+\frac{c_0c_1m_g^2}{2}r+c_0^2c_2m_g^2+\dot{R}\right)^\frac{1}{2}}{\displaystyle 1+\frac{ 8\pi P}{3}r^2-\frac{2M}{r}+\frac{q^2}{r^{2}}+\frac{c_0c_1m_g^2}{2}r+c_0^2c_2m_g^2},
\label{17}
\end{eqnarray}
and the internal of the shell denoted by $(M+dM,Q+dQ,P+dP,r+dr,c_1+dc_1,c_2+dc_2)$
\begin{equation}
n^{r-}=-\left(1-\frac{2(M+dM)}{r}+\frac{(Q+dQ)^2}{r^2}+\frac{8\pi (P+dP)}{3}r^2+\frac{c_0(c_1+dc_1)m_g^2}{2}r
+c_0^2(c_2+dc_2)m_g^2+\dot{R}^2\right)^\frac{1}{2}.
\end{equation}
Subsequently, when the shell goes near the horizon, the mass of the shell can be calculated via
\begin{eqnarray}
\label{mass}\mu&=&-R(n^{r+}-n^{r-})=-R
\left(1+\frac{8\pi P}{3}r^2-\frac{2M}{r}+\frac{Q^2}{r^{2}}+\frac{c_0c_1m_g^2}{2}r+c_0^2c_2m_g^2+\dot{R}^2\right)^\frac{1}{2}\\
&-&R\left(1-\frac{2(M+dM)}{r}+\frac{(Q+dQ)^2}{r^2}+\frac{8\pi (P+dP)}{3}r^2+\frac{c_0(c_1+dc_1)m_g^2}{2}r+c_0^2(c_2+dc_2)m_g^2+\dot{R}^2\right)^\frac{1}{2}\nonumber
\end{eqnarray}
where $dP$ depicts the variation of the shell's pressure. If we want to vary the value of $dc_1$ and $dc_2$, they can be seen as special charges labeled as $Q_{c_1}$ and $Q_{c_2}$, respectively. The terms with $c_i$ in the equation above can be treated as a special form of energy which is similar to the charge $Q$.

To make sure that the particle does not recoil outside the horizon and falls into the
black hole successfully, the absorption condition that the energy of the thin shell of dust must be greater than the minimum value should be satisfied.
 The minimum energy by using the condition $n^{r+}-n^{r-}\geq0$ can be acquired as
\begin{equation}
dM\geq \frac{q dQ}{r}+\frac{d^2Q}{2r^2},
\end{equation}
where the second term is the self-interaction of the particle which can be ignored.
In \eqref{mass}, we have $\dot{R}=dR/d\tau=u^r$, thus, we can solve out $M$ and do the variation as
\begin{equation}
dM=\frac{1}{2} c_0^2 m_g^2 r dc_2+\frac{1}{4} c_0 m_g^2 r^2 dc_1+\frac{4}{3} \pi r^3  dP+\frac{(dQ)^2}{2r}+\frac{ Q }{r}dQ-\frac{\mu ^2}{2 r}+\mu n^r .\label{dm}
\end{equation}
Here
\begin{equation}
n^r=\sqrt{1+c_0^2 c_2 m_g^2+\frac{1}{2}c_0 c_1 m_g^2 r-\frac{2 M}{r}+\frac{8}{3}\pi P r^2+\frac{Q^2}{r^2}+\dot{R}^2}.
\end{equation}
where $\mu$ and  $n^r$ depict the rest mass and the radial velocity of the shell, respectively. We take it as a positive value to ensure that the shell indeed arrives at the horizon of the black hole successfully. The higher order of $\mu$ and $dQ$ denote the self-interaction of particles which can be neglected in our work. So \eqref{mass} can be re-written as
\begin{equation}
dM=\frac{4}{3} \pi r^3  dP+\frac{ Q dQ }{r}+\frac{1}{4} c_0 m_g^2 r^2 dc_1+\frac{1}{2} c_0^2 m_g^2 r dc_2+\mu n^r . \label{dm2}
\end{equation}

Next up we will discuss the thermodynamics with the variation of mass calculated by the shell in the extended phase space. The increase in energy is related to the black hole's mass. So we can discuss the violation of thermodynamics with the expression of $dr_h$ which can be given by inserting \eqref{dm2} into \eqref{27}. Then we could delete $dM$ directly. Interestingly, $dQ,dc_1,dc_2$ can be deleted simultaneously, so we have the expression of $dr_h$ near the horizon
\begin{equation}
dr_h=\frac{12 \mu n^r r}{-12 M+r \left(12+12 c_0^2 c_2 m_g^2+9c_0 c_1 m_g^2r+64 \pi P r^2\right)}.\label{dr2}
\end{equation}
Using the relation between the entropy(volume) and radius with $dS=2\pi r_h dr_h$ and $dV=4/3 \pi r_h^2 dr_h$, yield $dS_h,dV_h$
\begin{eqnarray}
dS_h&=&\frac{24\pi \mu  n^r r^2}{-12 M+r \left(12+12 c_0^2 c_2 m_g^2+9c_0 c_1 m_g^2r+64 \pi P r^2\right)},\\ \label{ds2}
dV_h&=&\frac{16\pi \mu  n^r r^3}{-12 M+r \left(12+12 c_0^2 c_2 m_g^2+9c_0 c_1 m_g^2r+64 \pi P r^2\right)}.\label{ds2a}
\end{eqnarray}
Based on the formulas above, we obtain $dM=TdS_h+\Phi dQ+VdP+A dc_1+B dc_2$, which implies that the first law of thermodynamics is recovered by the dust falling.

We then discuss the validity of the second law via the entropy variation \eqref{ds2}. For the extremal black hole where $T=0$, substituting \eqref{35} into \eqref{ds2}, we can get $dS_h=\infty$.
For the non-extremal black hole, the event horizon should be larger than $r_h$. We plot the denominator of $dS_h$ the fixed $c_1=5,c_0=100, m_g=0.01, p=1£¬M=2, Q=0.5$ for different $c_2$, the conclusion is the denominator of $dS_h$ is always larger than zero which depicted in Fig.\ref{6}.

\begin{figure}
   \centering
\includegraphics[width=8cm]{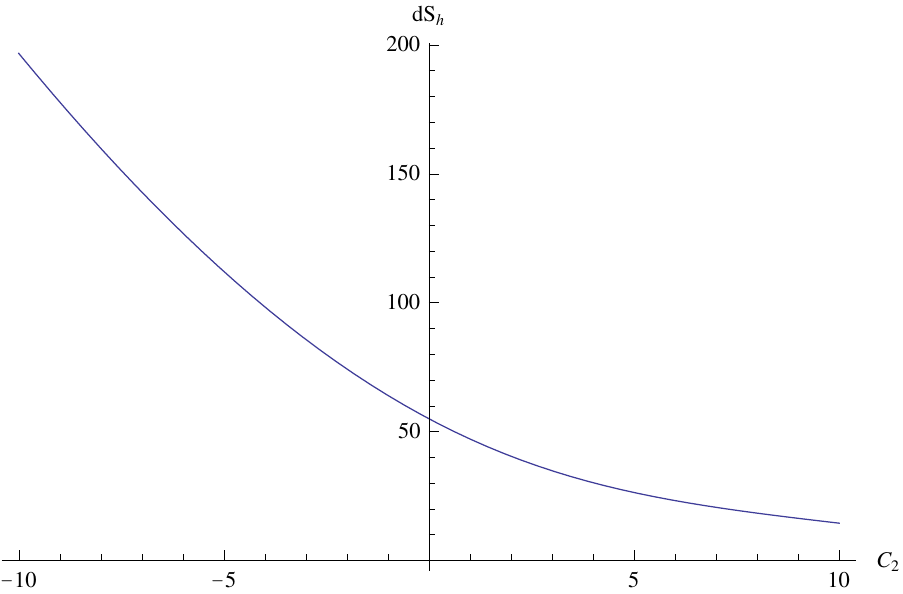}
\caption{ $dS_h$ for given $c_1=5$.}   \label{fig6}
\end{figure}

\begin{figure}
   \centering
\includegraphics[width=8cm]{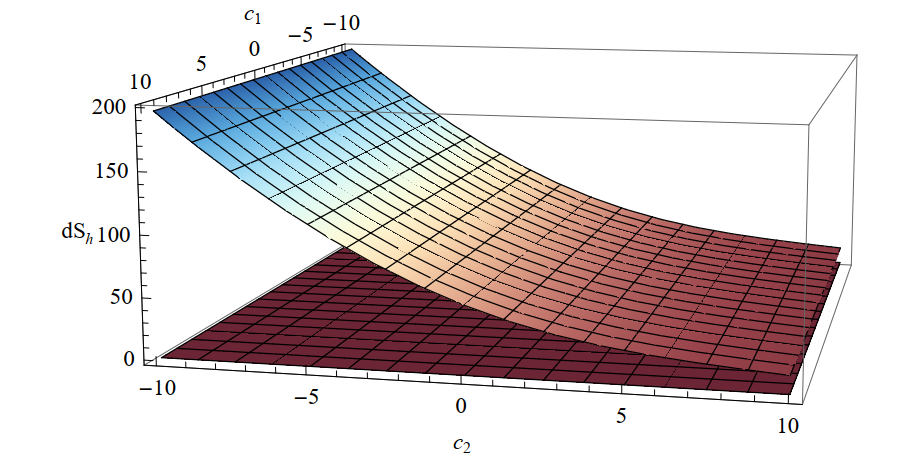}
\caption{The dependence of $dS_h$ on $c_1$ and $c_2$ after the dust falling.}   \label{fig7}
\end{figure}
We can see from the Fig.\ref{fig7} that $dS_h$ is always positive, no matter the value of $c_1$ and $c_2$, the entropy of the black hole will not decrease. This shows that the second law of thermodynamics is also valid.

What we discussed above is  in the extended phase space that with the dust falling method, the second law of thermodynamics is always valid. The conclusion is very different from that  in section \uppercase\expandafter{\romannumeral3},
where we assume that the black hole mass is regarded as an enthalpy which relates to the internal energy as $U=M-PV-A c_1-Bc_2$, the variations of entropy are determined by the value of $c_1$ and $c_2$.
This means that the results are very different in both two methods. However, in the normal phase space, the conclusion is very similar because the variation of energy is both related to the black hole's mass without $PV$ term. For simplicity, we don't show the detailed calculations here.

\section{Weak cosmic censorship conjecture}\label{sec:wcc}
In this section, we are going to briefly discuss the weak cosmic censorship conjecture under charged particle absorption. WCCC states that there is not a naked singularity for an observer at the future null infinity. It means that singularities should be hidden by the event horizon of  black hole. We will visit the variation under a charged particle absorbed by massive black hole to ensure the event horizon exists and the black hole is still not destroyed.
We mainly follow the strategy addressed in \cite{Gwak:2017kkt}.

There is always a minimum value for the function of $f(r)$ with a radial radius $r_m$. If the event horizons exist, the condition of $f(r_m)\leq 0$ should be satisfied. At $r_m$, we have
\begin{eqnarray}
f|_{r=r_m}\equiv f_m=\delta \leq 0,\nonumber\\
\partial_{r}f|_{r=r_m}\equiv f_m'=0,\label{38}\\
(\partial_{r})^2f|_{r=r_m}\equiv f_m''>0.\nonumber
\end{eqnarray}
For the extremal condition, we have $\delta=0$ is satisfied, otherwise, $\delta$ is a small
quantity. After the charged particle absorbed by the black hole, the minimum point becomes $r_m+dr_m$. Correspondingly, the variation of $f(r)$ at $r=r_m$ is
\begin{equation}
\partial_r f|_{r=r_m+dr_m}=f_m'+df_m'=0.\label{40}
\end{equation}
In the extended thermodynamics, we have
\begin{equation}
df_m'=\frac{\partial{f_m'}}{\partial M}dM+\frac{\partial{f_m'}}{\partial Q}dQ+\frac{\partial{f_m'}}{\partial P}dP+\frac{\partial{f_m'}}{\partial r_m}dr_m+\frac{\partial{f_m'}}{\partial c_1}dc_1+\frac{\partial{f_m'}}{\partial c_2}dc_2=0. \label{41}
\end{equation}
The physical parameters of the black hole change from the initial state $(M,Q,P,c_1,c_2,r_m)$ to the finial state described by $(M+dM,Q+dQ,P+dP,c_1+dc_1,c_2+dc_2,r_m+dr_m)$ which can be expressed as
\begin{equation}
f(r_m+dr_m)=f_m+df_m=\delta+\frac{\partial{f_m}}{\partial M}dM+\frac{\partial{f_m}}{\partial Q}dQ+\frac{\partial{f_m}}{\partial P}dP+\frac{\partial{f_m}}{\partial c_1}dc_1+\frac{\partial{f_m}}{\partial c_2}dc_2\label{43}
\end{equation}
where
\begin{equation}
\frac{\partial{f_m}}{\partial M}|_{r=r_m}=-\frac{2}{r_m},~~~
\frac{\partial{f_m}}{\partial Q}|_{r=r_m}=\frac{2Q}{r_m^2},~~~
\frac{\partial{f_m}}{\partial P}|_{r=r_m}=\frac{8\pi r_m^2}{3},~~~
\frac{\partial{f_m}}{\partial c_1}|_{r=r_m}=\frac{c_0 m_g^2r_m}{2},~~~
\frac{\partial{f_m}}{\partial c_2}|_{r=r_m}=c_0^2m_g^2.\label{partial}
\end{equation}

The key point is to analyze the sign of \eqref{43}.
Firstly, we discuss the case for the extremal black hole where we have $\delta=0$. We deform $dM$ with the condition of $f_m'=0$,
\begin{equation}
dM=\frac{\partial{M}}{\partial P}dP+\frac{\partial{M}}{\partial Q}dQ+\frac{\partial{M}}{\partial c_1}dc_1+\frac{\partial{M}}{\partial c_2}dc_2+\frac{\partial{M}}{\partial r_m}dr_m.\label{44}
\end{equation}
Combining \eqref{44} with \eqref{31}, $dr_m$ can be written as
\begin{equation}
dr_m=-\frac{r(-2Q dQ +2c_1 r dA+2 r c_2 dB+2 r |P^r|+c_0^2 m_g^2 r^2 dc_2+c_0 m_g^2 r^3 dc
_1+8\pi r^4 dP)}{2 Q^2+c_0 c_1 m_g^2 r^3+24 P\pi r^4}.\label{45}
\end{equation}
Then, inserting \eqref{44} into \eqref{41}, we obtain a conclusion of $dr_m=0$, and with the help of \eqref{45}, we have
\begin{equation}
dP=\frac{2Q dQ -r(2c_1 dA+2c_2 dB+2 |P^r|+c_0^2 m_g^2 r dc_2+c_0 m_g^2 r^2 dc
_1)}{8 P\pi r^4}.\label{46}
\end{equation}
 Substituting \eqref{46} and \eqref{45} into \eqref{43}, the transformation of $f(r_m+dr_m)$ is expressed as
\begin{equation}
f_m+df_m=-\frac{2 |P^r|}{r_m}.\label{48}
\end{equation}
The negative value after absorbing a charged particle implies that the final state of the extremal black hole becomes a non-extremal black hole. This means that the event horizon always exists, which depicts that the weak cosmic censorship conjecture is valid in the extended phase space for the extremal black hole.

For the near-extremal black hole, the minimum value of $\delta$ is a small quantity and it can be labeled as $f(M,P,Q,c_1,c_2,r_0)=\delta_\epsilon\ll1$. Besides, the value of $f'(r_m)$ is very close to zero. Thus, we use $r_m=r¡ª_0+\epsilon$ to ensure the near extremity because \eqref{31} can not be used directly, with $0<\epsilon\ll1$. Then, we expand \eqref{31} near the minimum point as

\begin{eqnarray}
dM&=&\frac{Q}{r_m}dQ+\frac{c_0^2 m_g^2 r_m}{2}  dc_2+\frac{c_0 m_g^2 r_m^2}{4}  dc_1 +\frac {4\pi r_m^3}{3} dP\nonumber\\
&+&\left (\frac{c_0^2 c_2 m_g^2 r_m}{2}  -\frac{Q^2}{2r_m^2}+\frac{c_0 c_1 m_g^2 r_m}{2} +4p\pi r_m^2 \right)dr_m\label{dm1}\\
&+&\left(\frac{c_0^2 m_g^2 }{2} dc_2+\frac{c_0 c_1 m_g^2}{2} dr_m+\frac{Q^2}{r_m^3}dr_m-\frac{Q}{r_m^2}dQ+\frac{c_0
 m_g^2 r_m}{2} dc_1+8 p\pi r_m dr_m+4\pi r_m^2 dP\right)\epsilon +\mathcal{O}(\epsilon)^2\nonumber.
\end{eqnarray}
Inserting \eqref{dm1} into \eqref{43}, we obtain
\begin{eqnarray}
df_m&=&c_0^2 m_g^2 dc_2+\frac{2Q}{r_m^2}dQ+\frac{1}{2} c_0 m_g^2 r_m dc_1+\frac83 \pi  r_m^2 dP
-\frac{dr_m}{r_m}-\frac{c_0^2 c_2  m_g^2}{r_m}  dr_m\nonumber\\
&+&\frac{ Q^2}{r_m^3}dr_m-\frac{2Q}{r_m^2}dQ
-c_0^2 m_g^2 dc_2-\frac{c_0 m_g^2 r_m}{2r_m}  dc_1-8 \pi P r_m dr_m-\frac{8\pi r_m^2}{3}  dP\label{dfmm}\\
&-&\frac{2}{r_m}(\frac{c_0 c_1  m_g^2}{2} dr_m+\frac{ c_0^2 m_g^2}{2} dc_2+\frac{ Q^2}{ r_m^3}dr_m-\frac{ Q}{r_m^2} dQ+\frac{c_0 m_g^2 r_m}{2}  dc_1+8 \pi P r_m dr_m+4 \pi r_m^2 dP)\epsilon+\mathcal{O}(\epsilon)^2.
\nonumber
\end{eqnarray}
We solve the expression of $P$ with $f'(r_h)=0$,
\begin{equation}
P=\frac{Q^2-r_m^2-c_0^2 m_g^2 r_m^2 c_2-c_0 c_1  m_g^2 r_m^3}{8\pi r_m^4}.\label{p}
\end{equation}
Naturally, the expression of $dP$ reads
\begin{eqnarray}
dP&=&\frac{Q}{4\pi r_m^4}-\frac{c_0^2 m_g^2 dc_2}{8\pi r_m^2}-\frac{c_0 m_g^2}{8\pi r_m}dc_1\nonumber\\
&+&\left(\frac{-2r_m-2c_0^2 c_2 m_g^2 r_m-3 c_0 c_1 m_g^2 r_m^2}{8\pi r_m^4}
-\frac{Q^2-r_m^2-c_0^2 c_2 m_g^2 r_m^2-c_0 c_1 m_g^2 r_m^3}{2\pi r_m^5}\right)dr_m. \label{dp}
\end{eqnarray}
Substituting \eqref{p} ,\eqref{dp} into \eqref{dfmm}, we have
\begin{equation}
df_m= \mathcal{O}(\epsilon)^2,
\end{equation}
such that
 \begin{equation}
 f_m+df_m=\delta_\epsilon+\mathcal{O}(\epsilon)^2.
 \end{equation}
Since $\delta_\epsilon$ and $\mathcal{O}(\epsilon)^2$ are small quantities which can be neglected directly. It implies the weak cosmic censorship is still valid for the near-extremal black hole.

Thus, under particle absorption, even though the second law of thermodynamics may violate in certain model parameters, the weak cosmic censorship is valid in the extended phase space.

We continue to check the weak cosmic censorship conjecture in normal phase space. To proceed, we write the changes in the conserved quantities of the black hole as $(M+dM,Q+dQ)$, and the locations of the minimum value and the event horizon can be written as $r_m+dr_m$, $r_h+dr_h$, respectively. Subsequently,
the variation of $f(r_m+dr_m)$ is
\begin{equation}
f(r_m+dr_m)=f_m+df_m=\delta+\frac{\partial{f_m}}{\partial M}dM+\frac{\partial{f_m}}{\partial Q}dQ.\label{f2}
\end{equation}
At $r_m+dr_m$, the variation of $f(r)$ is defined as $df_m$ and considering $f'_m=0$, we have
\begin{equation}
 df'_m=\frac{\partial{f'_m}}{\partial M}dM+\frac{\partial{f'_m}}{\partial Q}dQ+\frac{\partial{f'_m}}{\partial r_m}dr_m . \label{df2}
 \end{equation}
Also, with the condition $f'_m=0$, we obtain $M$ and further $dM$ as
\begin{equation}
dM=dR \left(-\frac{1}{2} c_0 c_1 m_g^2 r-8 \pi  P r^2-\frac{Q^2}{r^2}\right)+\frac{2 Q  }{r}dQ.\label{53}
\end{equation}
Inserting \eqref{53} into \eqref{df2}, we get
\begin{equation}
dR=\frac{2 r (|P^r|  r-QdQ )}{c_0 c_1 m_g^2 r^3+16 \pi  P r^4+2 Q^2}.\label{54}
\end{equation}
Combining \eqref{53} and \eqref{f2}, we obtain a conclusion of the minimum point
 \begin{equation}
 f_m+df_m=-\frac{2 |P^r|}{r}
 \end{equation}
which is the same as the result\eqref{48} for the extended phase space. We then will not repeat the
discussions in this case.

From the above analysis, we can conclude that under particle absorption, the weak cosmic censorship is valid both
in the extended phase space and in the normal phase space, even though the second law of thermodynamics may violate in certain model parameters. This means that the violation of the second law of thermodynamics under the assumption
does not affect weak cosmic censorship conjecture. Thus, we argue that the
assumption in the first method may be the reason for the violation of the second law of thermodynamics
and requires further careful treatment.

\section{Conclusion and discussion}
In this paper, we studied the thermodynamical laws of massive gravity black hole via two methods. First, we investigated the thermodynamics by considering the charged particle absorbed by the black hole and checked the validity of the thermodynamical laws both in the normal and extended phase space. In this method, the first law is saved, while the expression of the entropy could be negative depending on the model parameters and the second law of thermodynamics could be violated in extended thermodynamics in which  the cosmology constant is treated as the pressure of black hole. When it comes to the case with normal thermodynamics without $PV$ term, the violation does not appear.

Then we applied another method with a shell of dust falling into the massive gravity black hole, in which the mass of the shell can be calculated directly when it approaches the horizon of the black hole. We found the expressions of the horizon and entropy always larger than zero, which implies the first and second laws of thermodynamics are always valid in both cases with or without $PV$ term.

We argued that the violation of the second law in the extended thermodynamics may be brought by the assumption in the first method, which states that the particle changes the internal energy of the black hole. So the assumption should be carefully treated.

Finally, we checked the weak cosmic censorship conjecture in the first method. Our analysis shows that the weak cosmic censorship is always valid in both normal and extended phase space. This further supports our opinion that the assumption may contribute to the violation of the second law of thermodynamics.
It was argued in \cite{Hu:2019lcy} that under this assumption when the particle energy gives an increase in $M$, it will increase $V$ at the same time, cause the increase in the internal energy
is less than the increase in the enthalpy so that the assumption does not fit very well. Further details are called for.

\section*{Acknowledgements}
We appreciate De-Chang Dai,Yen Chin Ong and Jian-Pin Wu for helpful discussions. This work is supported by the Natural Science Foundation of China under Grant No.11705161, No.11675139 and
No.11875220.  Xiao-Mei Kuang is also partly supported Natural Science Foundation of Jiangsu Province under Grant No.BK20170481.  Shi-Qian Hu is also supported by the Postgraduate Reaearch \& Practice Innovation Program of Jiangsu Province (KYCX19\_2098).

\begin{appendices}
\section{Technique For Calculating Infalling Shell}
 In this appendix, we will show the technique of calculating the mass of shell, $\mu$, which is equal to the shift in the mass of black hole, $dM$  when the dust as a shell falling near the horizon as proposed in \cite{Alan P. Lightman:1975ab}. Considering that $\tau$ is a proper time measured by an observer at rest in the dust and the motion of the shell is specified by $r(\tau)$, then  the metric of the shell is specified by
\begin{equation}
ds^2=-d\tau^2+R^2(\tau)(d\theta^2+\sin^2{\theta}d\phi^2),
\end{equation}
where $R(\tau)=r(\tau)$ and  $4\pi R^2$ is the surface area of the shell at $\tau$.
We consider the equation of continuity
\begin{equation}
0=\frac{d\sigma}{d\tau}+\sigma u^i|_i=(\sigma u^i)|_i=(\sigma(^{(3)}g)^{\frac{1}{2}}u^i),_i/(^{(3)}g)^{\frac{1}{2}},
\end{equation}
where $\sigma$ is a surface mass density and $^{(3)}g$ is the determinant of the 3-dimensional metric. Together with $u^\tau=1$ and $^{(3)}g=R^4(\tau)$, we obtain $(\sigma R^2)_{,\tau}=0$ which implies that the rest mass of the shell is a constant with $4\pi R^2 \sigma=\mu$.

The discontinuity in extrinsic curvature $K$ is defined as $[K_{ij}]=8\pi\sigma(u_i u_j+\frac{1}{2}^{(3)}g_{ij})$ (please see \cite{Alan P. Lightman:1975ab} for more details),  where $u$ denotes the 4-velocity. To proceed, we calculate $[K_{\theta\theta}]$ component as
\begin{equation}
[K_{\theta\theta}]=8\pi\sigma(u_\theta u_\theta+\frac{1}{2}^{(3)}g_{\theta\theta})=4\pi \sigma^{(3)}g_{\theta\theta}=\mu.
\end{equation}
On the other hand, the component $K_{\theta\theta}$ is computed from the definition as
\begin{equation}
K_{\theta\theta}=-n_{\theta;\theta}=n_\alpha \Gamma_{\theta\theta}^\alpha=-\frac{1}{2}n^rg_{\theta\theta,r}=-r n^r.
\end{equation}
Then the discontinuity part can be derived as
\begin{equation}\label{DK}
[K_{\theta\theta}]=\mu=-r(n^{r+}-n^{r-})
\end{equation}
where $n^{r+}$ and $n^{r-}$ are the radial components of the normal evaluated in the exterior and the interior geometry.

When the shell fall into a black hole, we can calculate the external components $n^{r+}$ in the exterior geometry and internal components $n^{r-}$ in the interior geometry.
Specially, employing  $u.n=0$ and $n.n=-u.u=1$, we can evaluate the components of $u$ and $n$. Thus, exterior to the shell we have
\begin{eqnarray}
1&=&f(r)(u^t)^2-(u^r)^2/f(r),\\
0&=& n_r u^r + n_t u^t,\\
1&=&-f(r)^{-1}(n_t)^2+f(r)(n_r)^2.
\end{eqnarray}
After eliminating $u^t,n_t$ we obtain
\begin{equation}
n_r^+=\left[\frac{1+(u^r)^2/f(r)}{f(r)}\right]_{s+}^\frac{1}{2}.
\end{equation}
Moreover, one has $r=R(\tau)$ and $u^r=dR/d\tau\equiv\dot{R}$ on the shell, so  exterior to the shell, the contravariant component of $n$ is
\begin{equation}
n^{r+}\equiv g^{rr}n_r^+=\left(f(r)+\dot{R}^2\right)^\frac{1}{2}\mid_{s+}.
\end{equation}
With the same algebra, one can obtain $n^{r-}=\left(f(r)+\dot{R}^2\right)^\frac{1}{2}\mid_{s-}$. Therefore,  \eqref{DK} can be re-evaluated as
\begin{equation}
\mu=-R(n^{r+}-n^{r-})=-R\left[(f(r)+\dot{R}^2)^\frac{1}{2}\mid_{s+}-(f(r)+\dot{R}^2)^\frac{1}{2}|_{s-}\right],
\end{equation}
which is treated as $dM$ of the shift of the mass when the shell is near the horizon.
 \end{appendices}

 \end{document}